\documentstyle[12pt,epsfig,epsf,graphicx]{article}
\def\be{\begin{equation}}
\def\ee{\end{equation}}
\def\beq{\begin{eqnarray}}
\def\eeq{\end{eqnarray}}

\begin{document}
\begin{flushright}
TIFR/TH/99-42\\
KIAS/P/99079\\
DESY/99-132
\end{flushright}
\bigskip
\begin{center}
{\Large{\bf Signature of Charged to Neutral Higgs Boson Decay at the LHC
in SUSY Models}} \\[1cm]
Manuel Drees$^1$, Monoranjan Guchait$^2$ and D.P. Roy$^3$ 
\end{center}
\bigskip
\begin{enumerate}
\item[{}] $^1$IFT, Univ. Estadual Paulista, 01405-900, S\~ao Paulo,
Brazil 
\item[{}] $^2$DESY, Notkestrasse 85, D-22607 Hamburg, Germany
\item[{}] $^3$TIFR, Homi Bhabha Road, Mumbai - 400 005, India 
\end{enumerate}
\bigskip
\begin{center}
\underbar{\bf Abstract}
\end{center}
\medskip

We study the signature of $H^\pm$ decay into $h^0W$ at the LHC in SUSY
models.  It has only marginal viability in the MSSM.  But in the
singlet extensions like the NMSSM one can have a spectacular signature
for $H^\pm$ decay into $(h^0,A^0)W$ over a significant domain of
parameter space. 

\newpage

The minimal supersymmetric Standard Model (MSSM) contains two complex
Higgs doublets, $\phi_1$ and $\phi_2$, corresponding to eight scalar
states.  Three of these are absorbed as Goldstone bosons leaving five
physical states -- the two neutral scalars $(h^0,H^0)$, a
pseudo-scalar $(A^0)$ and a pair of charged Higgs bosons $(H^\pm)$.
All the tree-level masses and couplings of these particles are given
in terms of two parameters, $m_{H^\pm}$ and $\tan\beta$, the latter
representing the ratio of the two vacuum expectation values [1].
While any one of the above neutral Higgs bosons may be hard to
distinguish from that of the Standard Model, the $H^\pm$ carries a
distinctive signature of the SUSY Higgs sector.  Moreover the
couplings of the $H^\pm$ are uniquely related to $\tan\beta$, since
the physical charged Higgs boson corresponds to the combination 
\be
H^\pm = -\phi^\pm_1 \sin\beta + \phi_2^\pm \cos\beta.
\label{one}
\ee
Therefore the detection of $H^\pm$ and measurement of its mass and
couplings are expected to play a very important role in probing the
SUSY Higgs sector. 

So far the investigations of the $H^\pm$ signature have been based on
its couplings to fermionic channels, which constitute the dominant
decay channels of $H^\pm$ [2,3].  In this note we investigate the
$H^\pm$ signature at LHC in the bosonic decay channel [4]
\be
H^\pm \rightarrow Wh^0.
\label{two}
\ee
Although a subdominant channel for $H^\pm$ decay, it is important for
several reasons. 

\begin{enumerate}
\item[{i)}] It is the second most important channel in the low
$\tan\beta$ region for $m_H > m_t$, where the dominant decay channel
[2] 
\be
H^+ \rightarrow t\bar b
\label{three}
\ee
suffers from a large irreducible background from the QCD processes
\be
gg \rightarrow t\bar t g, ~gq \rightarrow t\bar tq, ~gb \rightarrow
t\bar tb.
\label{four}
\ee
\item[{ii)}] The fermionic couplings of $H^\pm$ hold for any two Higgs
doublet model of type II [1], i.e. where $H^\pm$ represents the
combination (\ref{one}).  In contrast the prediction of $H^\pm$
coupling to the $Wh^0$ channel in terms of $\tan\beta$ and $m_{H^\pm}$
holds only in the MSSM and hence serves as a unique test of this
model.
\item[{iii)}] Moreover, unlike its fermionic couplings the $H^\pm$
coupling to $Wh^0$ is sensitive to the singlet extensions of the
MSSM, like the next-to-minimal supersymmetric Standard Model [5,6]
(NMSSM). Therefore this signature would be useful in probing such
models. In fact we shall see below that the signature is expected to
have only marginal viability in the MSSM, while in NMSSM it can be
quite spectacular over a significant part of the parameter space.
\end{enumerate}

For calculating the branching fraction for $H^\pm \rightarrow Wh^0$
decay, we shall use the radiatively corrected MSSM relation between
$H^\pm$ and $h^0$ masses including the effect of stop mixing [7].  It
is only for very large values of the stop mass and mixing parameters
that the $h^0$ masses can escape the LEP-2 limit in the low
$\tan\beta$ region of our interest [8].  We shall therefore assume a
large stop mass of $\sim 1$ TeV along with maximal mixing.  The
relevant formulae for this decay branching fraction can be found in
the paper of Djouadi, Kalinowski and Zerwas [4].  The resulting
branching fractions are shown as functions of $\tan\beta$ in Fig. 1
for several $H^\pm$ masses.  In each case the region to the left of
the cross will be excluded by the nonobservation of $h^0$ at LEP-2 at
the end of the 200 GeV run.  In fact the present exclusion limits of
LEP-2 are close to these values. 

Fig. 1 shows that the $H^\pm \rightarrow Wh^0$ decay branching
fraction is at best $8-9$\% for $m_{H^{\pm}} = 200$ GeV, i.e. just above the
$t\bar b$ threshold, and goes down rapidly with increasing $m_{H^\pm}$.
Note that even below the $t\bar b$ threshold (dashed line) this
branching fraction 
remains $\leq 4-5$\% over the LEP-2 allowed region.  The latter
corresponds to $m_h > 100$ GeV, which means that $m_{H^\pm}$ is also below
the $Wh^0$ threshold.  The dominant decay channel in this case is
$H^\pm \rightarrow \tau \nu$ [3,4].  We shall see below however that
the LEP-2 constraints on the $h^0,A^0$ masses are far less severe in
the NMSSM.  Consequently a $H^\pm$ below the $t\bar b$ threshold can
decay into on-shell $Wh^0 (A^0)$, which would then be the dominant
decay channel.

For $m_H > m_t$ the dominant process of $H^\pm$ production at LHC is
its associated production with top.  We shall be using the LO
production mechanism
\be
gb \rightarrow tH^- + {\rm h.c.}
\label{five}
\ee
for computing the signal cross-section.  
It is controlled by the
Yukawa coupling at the $tbH$ vertex,
\be
{g \over \sqrt{2} m_W} H^+ \left[\cot\beta m_t \bar t b_L + \tan \beta
m_b \bar t b_R\right] + {\rm h.c.}
\label{six}
\ee
Consequently it goes down like $1/\tan^2\beta$ over the low
$\tan\beta$ region of our interest.  The electro-weak loop correction
to this cross-section has been estimated to give upto 20\% reduction
in this region [9].  The corresponding QCD correction is expected to
be larger, but not yet available.  It may be noted here that the SUSY
QCD correction to the $H^\pm$ signal cross-section could be large and
of either sign depending on the choice of SUSY parameters [10].  For
simplicity we shall neglect this by assuming a large SUSY mass scale
$\sim 1$ TeV, which is consistent with our assumption of stop mass. 
Finally, the cross-section from $gg\rightarrow tbH^-$ is about half that of the
LO process (\ref{five}). Adding the two contributions and subtracting out the
overlapping piece seems to give a net signal cross-section midway between
the two values, i.e. about 2/3rd the LO Cross-section [11].
We shall be neglecting this correction to the LO cross-section as well.

Our analysis is based on a parton level Monte Carlo program.  The LO
cross-section for (\ref{five}) is convoluted with the LO parton
densities of CTEQ-4L [12] to generate the $tH^-$ signal.  This is
followed by the decay sequence
\be
tH^- \rightarrow bWh^0W \rightarrow bb\bar b \ell\nu q\bar q,
\label{seven}
\ee
i.e. $h^0 \rightarrow b\bar b$ while one $W$ decays leptonically and
the other hadronically.  Thus the final state consists of 3 $b$-tagged
and 2 untagged jets along with a hard lepton and missing $p_T$
$({p\!\!\!/}_T)$. We consider the background coming from the dominant
decay channel (\ref{three}) as well as the QCD processes
(\ref{four}). We have tried to simulate detector resolution by a
gaussian smearing of all the jet momenta, with 
\be
(\sigma(p_T)/p_T)^2 = (0.6/\sqrt{p_T})^2 + (0.04)^2.
\label{eight}
\ee
The ${p\!\!\!/}_T$ is obtained by vector addition of all the $p_T$'s
after resolution smearing. 

As a basic set of selection cuts we require
\be
p_T > 30 ~{\rm GeV} ~~~{\rm and}~~~ |\eta| < 2.5
\label{nine}
\ee
for all the jets and the lepton, where $\eta$ denotes pseudorapidity
and the $p_T$-cut is applied to ${p\!\!\!/}_T$ as well.  We also
require a minimum separation of 
\be
\Delta R = \left[(\Delta \phi)^2 + (\Delta \eta)^2\right]^{1/2} > 0.4
\label{ten}
\ee between the lepton and the jets as well as each pair of jets. We
require that exactly three $b-$quarks in the final state (7) are tagged,
assuming a $b$-tagging efficiency of 50\%. We take a mis-tagging
probability of 1\% for light quark and gluon jets when estimating the
signal and background cross-sections.

Then the signal and background events are subjected to the following
mass reconstructions. 
\begin{enumerate}
\item[{(a)}] The invariant mass of the two untagged jets is required
to be consistent with $m_W \pm 15$ GeV, and the resulting $W$ is
reconstructed from their momenta.
\item[{(b)}] For the leptonically decaying $W$ the $p_L(\nu)$ and the
resulting $p_L(W)$ are determined within a quadratic ambiguity using
the constraint $m_W = m(\ell\nu)$.  In the case of complex solutions
the imaginary part is discarded and the solutions coalesce.
\item[{(c)}] One of the reconstructed $W$'s is required to combine
with one of the $b$'s to give an invariant mass $= m_t \pm 25$ GeV.
In case of several combinations satisfying this constraint, the one
closest to $m_t (= 175~{\rm GeV})$ is selected.  The corresponding $b$
and $W$ are identified with the reconstructed top. 
\item[{(d)}] The remaining $b$-pair is required to have an invariant
mass $= m_h \pm 10$ GeV.  This constraint helps to suppress the
backgrounds from (\ref{three}) and (\ref{four}).
\item[{(e)}] Moreover we require that the invariant mass of the
remaining $W$ with neither of this $b$-pair should lie within $m_t \pm
20$ GeV.  This veto on the second top mass helps to suppress the
backgrounds further. 
\item[{(f)}] Finally the $H^\pm$ mass is reconstructed by combining
this $W$ with the $b$-pair.  In case this $W$ has a quadratic
ambiguity both the invariant masses are retained.  Of course only one
of these will correspond to the actual $H^\pm$ mass.
\end{enumerate}

Fig. 2 shows the signal and background cross-sections against this
$Wbb$ invariant mass for $m_{H^\pm} = 220$ GeV and $\tan\beta = 2$ --
i.e. for a $H^\pm$ mass just above the $t\bar b$ threshold and
$\tan\beta$ just above the corresponding exclusion range of LEP-2.
The mass constraints have helped to suppress both the backgrounds by
at least an order of magnitude each without any serious loss to the
signal cross-section.  Consequently the signal/background ratio is
$\sim 1$ in the region of the $H^\pm$ mass peak in the signal.
Unfortunately the signal size is rather marginal.  It corresponds to a
cross-section of $.04$ fb -- i.e. about a dozen events for an
accumulated luminosity of 300 fb$^{-1}$, corresponding to 3 years of
LHC run at high luminosity.  It should be mentioned here that this
choice of $H^\pm$ mass and $\tan\beta$ represents by far the most
favourable combination for the $H^\pm \rightarrow Wh^0$ signal (see
Fig. 1).  Increasing the $H^\pm$ mass from 220 to 300 GeV reduces the
decay branching fraction by a factor of 3 and the overall signal size
by a factor of $\sim 5$.  While there is a small increase in the
branching fraction with $\tan\beta$, it would be more than offset by
the corresponding drop in the $H^\pm$ production cross-section
(\ref{five}). 

We have also investigated the signal for $m_{H^\pm} = 160$ GeV at $\tan\beta
= 3$, i.e. at the edge of the LEP-2 exclusion range.  Here the main
contribution to the signal comes from 
\be
\bar tt \rightarrow \bar t b H^+,
\label{eleven}
\ee
followed by the decay chain (\ref{seven}).  The additional $b$ in
(\ref{eleven}) is found to be too soft to survive the $p_T > 30$ GeV
cut.  Thus the final state is practically the same as in
(\ref{seven}), except that one of the two $W$'s is off-shell.
Nontheless it is possible to do a complete reconstruction by requiring
leptonic decay for the on-shell $W$.  The reconstructed masses are
then subjected to the $m_t$ and $m_h$ constraints as before.  The
resulting signal is again found to be of similar size as in Fig. 2,
i.e. only $\sim .05$ fb.  The main reason for this is that the $t
\rightarrow bH^+$ branching fraction in this case is only 0.4\%.  Thus
the $H^\pm \rightarrow Wh^0$ decay signal is expected to be of only
marginal size, and that too over a limited range of the MSSM
parameters, $m_{H^\pm}$ and $\tan\beta$.

One can have a more favourable signal in singlet extensions of the
MSSM, where the LEP constraints on the Higgs boson masses are far less
severe.  Thus it is possible to have a $H^\pm$ lighter than top for
any value of $\tan\beta$ down to 1.5; and it is possible for this
$H^\pm$ to decay into a on-shell $Wh^0$ and/or $WA^0$ pair [13].  In
this case $H^\pm \rightarrow Wh^0 (A^0)$ is expected to be the
dominant decay channel, resulting in a spectacular signal at LHC.

Two types of singlet extensions of the MSSM have been extensively
discussed in the literature.  The first is based on a $U(1)$ extension
of the SM gauge group, which is inspired by $E(6)$ GUT [14,15].  The
second only extends the Higgs sector of the MSSM by adding a complex
singlet superfield $N$.  This is the above mentioned NMSSM, which is
widely recognised as offering a natural solution to the so-called
$\mu$-problem of the MSSM [5,6].  We shall be concentrating on this
second model. 

In the NMSSM the Higgs self-interaction is described by two cubic
terms in the superpotential, i.e.
\be
\lambda N H_1 H_2 - {k \over 3} N^3
\label{twelve}
\ee 
using the notation of [6]. Together with the corresponding soft
breaking terms, $A_\lambda$ and $A_k$, and the singlet vev $\langle N
\rangle$, there are 5 free parameters in addition to $\tan\beta$.
Thus the Higgs sector is less constrained than that of the MSSM, where
we had only one other free parameter $(m_{H^\pm})$ along with
$\tan\beta$. Consequently the MSSM mass relations among the physical
Higgs particles and the resulting indirect mass limits from LEP are no
longer valid. Moreover we have to add a singlet scalar and a
pseudoscalar, which will mix with the corresponding doublet states
diluting the direct mass bound on the latter from LEP.

We have numerically scanned the above 5 parameters to obtain solutions
which give maximal branching fractions for
\be
H^\pm \rightarrow W (h^0_1,A^0_1)
\label{thirteen}
\ee
for fixed input values of $\tan\beta$ and fixed output ranges of
$m_{H^\pm}$ [13]. Here the subscript 1 denotes the lightest scalar
(pseudoscalar) state. The radiative correction has been included
assuming a large stop mass of $\sim 1$ TeV and maximal stop mixing as
in the earlier case. We have included the final exclusion limits of
LEP-2 along with those from LEP-1.  Moreover we have also required
that the desired minimum of the Higgs potential, where all the three
neutral Higgs fields have non-zero vev, is the absolute minimum of the
potential.  This physical requirement helps considerably in
constraining the parameter space.

Table I shows the optimal solutions with $M_{H^\pm} \sim 160$ GeV for fixed
values of $\tan\beta = 2,2.5$ and $3$.  It should be noted that these
solutions are obtained with reasonable values of the singlet vev and
coupling parameters.  The resulting $h_1$ and $A_1$ masses are shown
along with the corresponding branching fractions.  For each
$\tan\beta$, we show two solutions where $H^\pm \rightarrow Wh_1$ and
$H^\pm \rightarrow WA_1$ are the dominant decay channels.  Note that
for dominant $H^\pm \rightarrow Wh_1$ solutions the $h_1$ mass is
always close to the LEP-1 bound from the $Z^*h$ final state.  Thanks
to the large event rate, the LEP-1 limit on $m_h$ is far more robust
than the corresponding limit from LEP-2.  Any $h^0$ with a
non-negligible doublet component cannot lie much below the former
limit. In contrast there is no direct limit on $A^0$ from LEP in the
low $\tan\beta$ region.  We have only associated production of $h^0
A^0$; and even this is strongly suppressed in the low $\tan\beta$
region.  There is an indirect limit obtained from the LEP limit on
$m_h$ using the MSSM mass relation, which is not valid here.
Consequently even a doublet pseudoscalar can be very light in this
model.  This explains why the optimal $H^\pm \rightarrow WA^0_1$
solutions favour so low values of $A_1^0$ mass.  Indeed this may be
the most promising process for $A^0$ search in the low $\tan\beta$
region.

We have estimated the signal cross-section from $t\bar t$ production,
followed by the decays (\ref{eleven}) and (\ref{seven}).  As mentioned
before, the accompanying $b$ in (\ref{eleven}) is too soft to survive
the $p_T > 30$ GeV cut, so that the final state is practically the
same as (\ref{seven}), with both $W$'s on-shell.  We apply the same
selection cuts and invariant mass constraints as before.  The
resulting signal cross-sections are listed in the last column.  We
have not listed the signal size for the light $A_1$ solutions, because
the small value of $m_{A_1}$ makes it very sensitive to the choice of
$\Delta R$ (\ref{ten}).  Of course such small values of $m_{A_1}$ are
picked up because of the requirement of maximal $B_{A_1}$, which is the  
branching fraction of $H^\pm$ to $W A_1$, in our
optimization program.  One can easily raise $m_{A_1}$ to $\sim 50$ GeV
without much reduction to the resulting $B_{A_1}$.  The resulting
signal would then be of similar size as the listed ones.

The signal size is $\sim 2$ fb, corresponding to about 200 events for
an annual luminosity 100 fb$^{-1}$ at LHC.  Evidently these would be
very spectacular events, consisting of 3 $b$'s and 2 $W$'s.  While one
of the $W$'s should combine with one of the $b$'s to form the $m_t$
peak, the remaining $b$-pair should show the $m_h (m_A)$ peak and also
combine with the remaining $W$ to form the $m_{H^\pm}$ peak.  We have also
checked that one can get similar solutions for still smaller values of
$m_{H^\pm}$ $(= 140-150~{\rm GeV})$ as well as $\tan\beta (= 1.5)$, which
would correspond to still larger signals.

\vspace*{5mm} \noindent
{\bf Acknowledgements:} This work was started at the Les
Houches Workshop on Physics at TeV colliders, organised by LAPP,
Annecy.  We thank the organisers, Patrick Aurenche and Fawzi Boudjema,
for a very stimulating workshop. 
The authors are thankful to Peter Zerwas and Michael Spira for useful
suggestions and Rajeev Bhalerao for computational advice.
The work of MG was supported by the Alexander
von Humboldt Fellowship while that of DPR was partly supported by the
IFCPAR under project No. 1701-1. MD thanks the School of Physics of
KIAS, Seoul, for their hospitality while this work was completed.


\begin{center}
{\bf References}
\end{center}
\bigskip

\begin{enumerate}
\item[{[1]}] J.F. Gunion, H.E. Haber, G.L. Kane and S. Dawson, ``The
Higgs Hunters' Guide'' (Addison-Wesley, Reading, MA, 1990).
\item[{[2]}] J.F. Gunion, Phys. Lett. B322 (1994) 125; V. Barger,
R.J.N. Phillips and D.P. Roy, Phys. Lett. B324 (1994) 236;
D.H. Miller, S. Moretti, D.P. Roy and W.J. Stirling, hep-ph/9906230.
\item[{[3]}] S. Raychaudhuri and D.P. Roy, Phys. Rev. D52 (1995) 1556;
D53 (1996) 4902; M. Guchait and D.P. Roy, Phys. Rev. D55 (1997) 7263;
E. Keith, E. Ma and D.P. Roy, Phys. Rev. D56 (1997) R5306; E. Ma,
D.P. Roy and J. Wudka, Phys. Rev. Lett. 80 (1998) 1162; D.P. Roy,
Phys. Lett. B459 (1999) 607.
\item[{[4]}] S. Moretti and W.J. Stirling, Phys. Lett. B347 (1995)
291, Erratum {\textit {ibid}}, B366 (1996) 451; A. Djouadi, 
J. Kalinowski and P.M. Zerwas, Z. Phys. C70 (1996)
435. 
\item[{[5]}] H.P. Nilles, M. Srednicki and D. Wyler, Phys. Lett. 120B
(1983) 346; M. Drees, Int. J. Mod. Phys. A4 (1989) 3635; 
J.Ellis, J.F.Gunion, H.E.Haber, L.Roszkowski and F.Zwirner,
Phys. Rev. D39 (1989) 844;
U. Ellwanger
and M. Rausch de Traubenberg, Z. Phys. C53 (1992) 521; P.N. Pandita,
Z. Phys. C59 (1993) 575; Phys. Lett. B318 (1993) 338; T. Elliot,
S.F. King and P.L. White, Phys. Rev. D49 (1994) 2435.
\item[{[6]}] S.F. King and P.L. White, Phys. Rev. D52 (1995) 4183; D53
(1996) 4049.
\item[{[7]}] H.E. Haber, R. Hempfling and A.H. Hoang, Z. Phys. C75
(1997) 539. M. Carena, J. Espinosa, M. Quiros and C. Wagner, Phys.
Lett. B355 (1995) 209. 
\item[{[8]}] see e.g. ALEPH Collaboration: R. Barate et. al.,
Phys. Lett. B440 (1998) 419. 
\item[{[9]}] L.G. Jin, C.S. Li, R.J. Oakes and S.H. Zhu,
hep-ph/9907482.
\item[{[10]}] J.A. Coarasa, D. Garcia, J. Guasch, R.A. Jimenez and
J. Sola, Eur. Phys. J. C2 (1998) 373; Phys. Lett. B425 (1998) 329.
\item[{[11]}] F. Borzumati, J. L. Kneur and Nir Polonsky, hep-ph/9905443.
\item[{[12]}] CTEQ Collaboration: H.L. Lai et. al., Phys. Rev. D55
(1997) 1280.
\item[{[13]}] M. Drees, E. Ma, P.N. Pandita, D.P. Roy and S. Vempati,
Phys. Lett. B433 (1998) 346.
\item[{[14]}] For a review, see J.L. Hewett and T.G. Rizzo,
Phys. Rep. 183 (1989) 193.
\item[{[15]}] H.E. Haber and M. Sher, Phys. Rev. D35 (1987) 2206;
M. Drees, Phys. Rev. D35 (1987) 2910; V. Barger and K. Whisnant,
Int. J. Mod. Phys. A3 (1988) 1907; E. Keith and E. Ma, Phys. Rev. D54
(1996) 3587; D56 (1997) 7155.
\end{enumerate}

\newpage
\[
\begin{tabular}{|c|c|c|c|c|c|c|c|c|}
\hline
$\tan\beta$ & $M_{H^\pm}$ & $B_{H^\pm}$ & $\langle N\rangle$ &
$\lambda, k$ & $A_\lambda,A_k$ & $m_{h_1},m_{A_1}$ & $B_{h_1},B_{A_1}$ &
$\sigma_{H^\pm}$ \\
& (GeV) & (\%) & (GeV) & & (GeV) & (GeV) & (\%) & (fb) \\
\hline
&&&&&&&& \\
 & 164 & 0.4 & 147 & .39,-.25 & -158,-59 & 56,36 & 51,43 & 2 \\
2&     &     &     &          &          &       &       & \\
 & 160 & 0.8 & 273 & .40,-.73 & 12, 8    & 115,15& 0,97  & -- \\
&&&&&&&& \\
\hline
&&&&&&&& \\
 &     &     & 231 & .21,-.41 & -101,111 & 51,137& 86,0  & 2.2 \\
2.5 & 160 & 0.5&   &          &          &       &       &     \\
 &     &     & 278 & .33,-.72 & 16,8     & 113,15& 0,95  & --  \\    
&&&&&&&& \\
\hline
&&&&&&&& \\
 &    &    &   196 & .14,-.33 & -184,-8 & 54,27 & 69,16 & 1.6 \\
3 & 160 & 0.4 & & & & & & \\
 &      &     & 341 & .22,-.62 & 23, 6 & 110,19 & 0,90 & -- \\
&&&&&&&& \\
\hline
\end{tabular}
\]
\begin{enumerate}
\item[{}] Table I - Maximal branching fractions for $H^\pm \rightarrow
W(h^0_1,A^0_1)$ decay in the NMSSM for fixed input values of $\tan\beta$
and output $H^\pm$ mass of $\sim 160$ GeV.  The values of the
$h^0_1,A^0_1$ masses and these branching fractions are shown along with the
corresponding model parameters.  Also shown are the $t \rightarrow b
H^\pm$ branching fraction and the size of the resulting $H^\pm
\rightarrow W(h^0_1,A^0_1)$ decay signal at LHC.
\end{enumerate}



\begin{figure}[htb]
\centerline{
\epsfysize = 1.0\textwidth 
\epsfxsize = 1.0\textwidth 
\epsffile{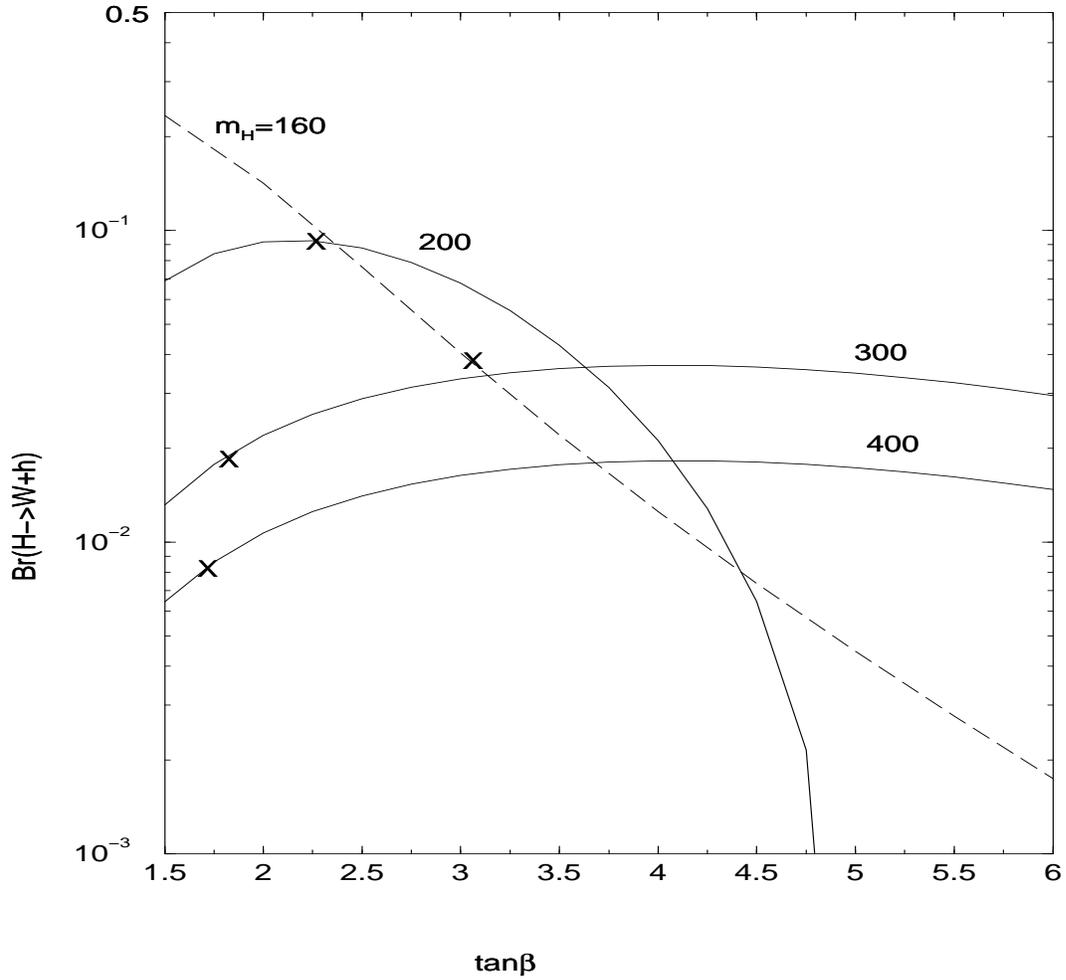}}

\vspace*{1.4cm}
\caption
{Branching fraction of $H^\pm \rightarrow Wh^0$ decay
is shown against $\tan\beta$ for different $H^\pm$ masses.  In each
case the LEP-2 exclusion limit of $\tan\beta$ is indicated by the
cross.} 
\end{figure}
\begin{figure}
\vspace*{-10.0cm}
\centerline{
\epsfxsize = 1.0\textwidth \epsffile{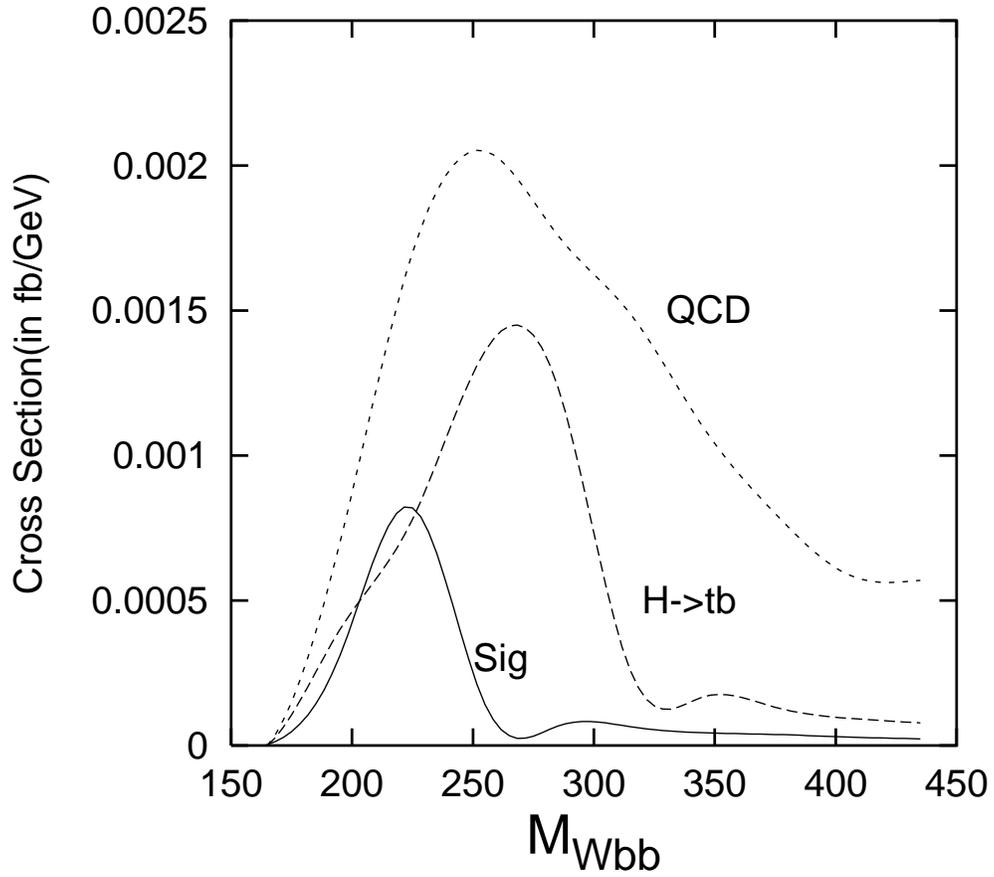}}

\caption{
The $H^\pm \rightarrow Wh^0$ signal cross-section at
LHC is shown against the reconstructed $H^\pm$ mass for $m_{H^\pm} = 220$
GeV and $\tan\beta = 2$ along with $H^\pm \rightarrow t\bar b$ and the
QCD backgrounds.}
\end{figure}

\end{document}